\documentclass{JAC2000}

\usepackage[dvips]{graphicx,rotating}
\usepackage{float}

\newcommand\hbt{high-$\beta$ }

\begin{document}   

\title{ PRECISE MEASUREMENT OF THE TOTAL CROSS SECTION AND  THE \\
        COULOMB SCATTERING AT THE LHC\thanks{Work partially 
        supported by contracts GVDOC99-01-8 and AEN98-1665-E}}

\author{ A. Faus-Golfe, J. Velasco\\  
    Instituto de Fisica Corpuscular CSIC - Universidad de Valencia\\
         M. Haguenauer\\  
    EP CERN - Ecole Polytechnique, France}

\maketitle   

\begin{abstract}     
                       A precise measurement of the total cross section
                       and  the Coulomb scattering at LHC requires 
                       the observation of elastically 
                       scattered particles at extremely small angles
                       (14 $\mu$rad, $-t \leq$ 0.01 GeV$^{2}$ for 
                       the first case; 3 $\mu$rad, $-t \leq$ 0.0006 GeV$^{2}$)
                       for the second one).
                       In this paper a very \hbt insertion optics which 
                       fulfills both conditions is presented.
                       A feasibility study, including the acceptance 
                       of the detectors, for an experiment to be installed
                       in IR1 or IR5, is also presented. 

\end{abstract}

\section{A PRECISE  MEASUREMENT OF TOTAL CROSS SECTION}
To determine $\sigma_{tot}$ with good precision ($ \le 2 \%$)
we must reach  $-t \simeq$ 0.01 GeV$^{2}$  with an acceptance better than
$50 \%$.
\subsection{Improving the standard layout: its limits}
From the acceptance studies \cite{accep}
of the standard \hbt optics described in 
\cite{ttm15-v6.0} we could conclude that to have more than 50 \% of
acceptance  for $-t$ = 0.01 GeV$^2$ we need to improve the optics, i.e.
we need to increase the  effective distance, $L_{z,eff}$ where $z=x,y$.
If we have a parallel to point focusing optics,
$(\phi_{z_d} - \phi^{*}_{z})= \pi/2,\ 3\pi/2,...$
the effective distance is given by:
$L_{z,eff} = M_{z,12} = \sqrt{ \beta^{*}_{z} \beta_{z_d} } $
Therefore the way to increase the effective distance for a 
given place, keeping the parallel to point focusing condition, 
is going far away from the IP.
This is  due to the fact that once you have fixed a place
and the phase advance the effective distance keeps constant.
This behavior is discussed in detail in \cite{beta-leff}.

Table \ref{ipd-ir5v6-1100} summarizes the performances
of various \hbt optics with $\beta^{*}$=1100 m
and different Roman pots positions, for Ring 1 calculated 
with MAD9 \cite{mad9}. 
A slight improvement is observed for  $M_{y,12_{d}}$
when we increase the distance
between the IP and the Roman pots.
 
\begin{table}[htb]
\centering
\begin{tabular}{| l | c | c | c | c |} \hline
optics  & \multicolumn{1}{c|}{ standard} & \multicolumn{2}{c|}{ new} & \multicolumn{1}{c|}{} \\ \hline
$\epsilon_{z}$    & \multicolumn{3}{c|}{$5.03\ \ 10^{-10}$} & m rad  \\ \hline
 
$\beta_{z}^{*}$    & \multicolumn{3}{c|}{1100.0} & m  \\  \hline

   \multicolumn{5}{|c|}{measurement vertical}   \\ \hline
   &  \multicolumn{2}{c|}{before $D2$} &
                  \multicolumn{1}{c|}{$Q4-Q5$} &
                  \multicolumn{1}{c|}{} \\ \hline
               &  \multicolumn{1}{c|}{}
               &  \multicolumn{1}{c|}{close $D2$} & 
                  \multicolumn{1}{c|}{close $Q4$} &
                  \multicolumn{1}{c|}{} \\ \hline
$\beta_{y_d}$         & 20.1  &  20.6  &   22.2   &m  \\  
$\Delta \mu_{y_d}$    &  0.250&   0.250&    0.250 & 2$\pi$  \\ 
$M_{y,12_{d}}$        &148.6  & 150.5  &  156.3   &m  \\
$|\theta_{y_{min}}|$  & 14.3  &  14.1  &   13.6   &$\mu$rad \\ 
$|t_{y_{min}}|$       &  0.010&   0.010&    0.009 &GeV$^2$  \\ \hline

\end{tabular}
\caption{ Performance of a total cross section  experiment 
        at the IP and at the detector place  of Ring 1
        for optics with $\beta^{*}$=1100, 
        Version 6.0 at 7 TeV for nominal emittance and 
        for different positions of 
        the Roman Pots ($RP2$ and $RP3$ in middle part of figure \ref{ir1}). 
 $|\theta_{y_{min}}| = \sqrt{2} y_{d} / M_{y,12_{d}}$ with $y_{d}$ = 1.5 mm.}
\label{ipd-ir5v6-1100}
\end{table}

We could conclude from the table that to improve drastically the 
$L_{z,eff}$ is necessary to go far away from the IP. This implies 
changes in the standard layout.

\section{A PRECISE  MEASUREMENT OF COULOMB SCATTERING}
The measurement of Coulomb scattering at LHC is important 
for at least two reasons: to use the coulomb
amplitude for normalization of $d\sigma/dt$ nuclear and to determine the real
part of the forward elastic scattering amplitude. The technique is
through the interference between the nuclear and coulomb amplitudes  whose
maximum is reached at $-t_{0} \simeq  8 \pi \alpha/\sigma_{tot}$, where
$\sigma_{tot}$ is the total cross section for
hadronic $p-p$ interactions.  At $\sqrt s \ = \ 14$ TeV, with $\sigma_{tot}$ 
predicted to be 110 mb \cite{UA42}, 
$-t_{0} \simeq 6 \times \ 10^{-4}$  GeV $^{2}$.  Scattering angles,
$\theta \simeq \sqrt{-t}/p$,   are of the order of  3 $\mu rad$. 
These angles are smaller that the typical
angular divergence of the beam in high luminosity operation, which is
$\Delta \theta = \sqrt{ \epsilon /\beta^{*} }\ge 35 \mu rad$.
\subsection{Requirements for the insertion optics}
Previously an optics for the measurement of total cross section
(14 $\mu$rad, $-t \leq$ 0.01 GeV$^2$)
was found \cite{ttm15} and \cite{ttm15-v6.0}.
Scaling it down to the requirements of Coulomb
scattering (3 $\mu$rad, $-t \leq$ 0.0006 GeV$^2$)
at the LHC energies 
gives us $M_{z,12}= L_{z,eff} \geq $ 707.6 m  taking a minimum
approach distance  of $\pm$1.5 mm.

\section{THE OPTICS}

\subsection{Hardware requirements for  very \hbt optics}

\begin{figure}[htb]
\centering
\hspace*{-1.5cm}
\includegraphics*[width=110.0mm]{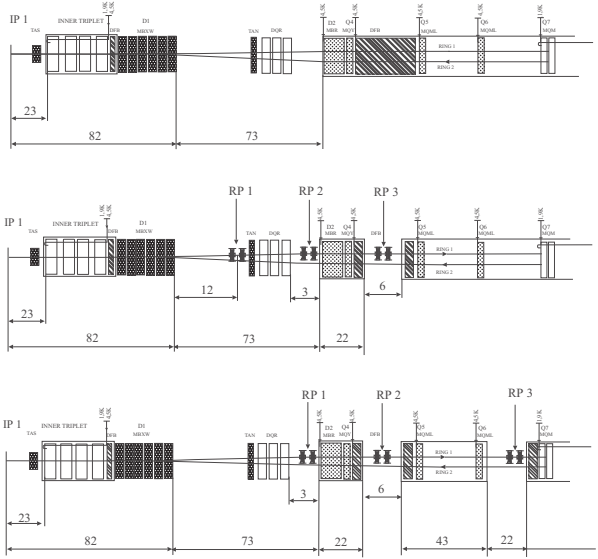}
\caption{Layout of the insertion 1, Version 6.0. The upper part reproduces the 
standard insertion, the middle part shows the location of the three
Roman pot stations $RP1$, $RP2$ and $RP3$ for the 
standard measurement of total
cross section while the lower part shows the location of the three
Roman pot stations $RP1$, $RP2$ and $RP3$ for a precise 
measurement of total cross section and Coulomb
scattering. The layout is symmetric with respect to the IP.}
\label{ir1}
\end{figure}

The layout of the right part of IR1 is shown in figure \ref{ir1}.
There are in principle four possible locations for the detectors, one 
just before the dipole $D2$, the second one between  $Q4$ and  $Q5$,
the third one between $Q5$ and $Q6$ and finally between $Q6$ and $Q7$.
In these two last cases a warm section would have to be provided.

\subsection{Optics solution for very \hbt optics}

Assuming the standard conditions of Version 6.0 described in 
\cite{ttm15-v6.0}
no solution could be matched which fulfills the requirements of parallel 
to point focusing optics.

\begin{figure}[htb]
  \begin{center}
    \leavevmode
\begin{turn}{-90}
     \includegraphics*[width=60.0mm]{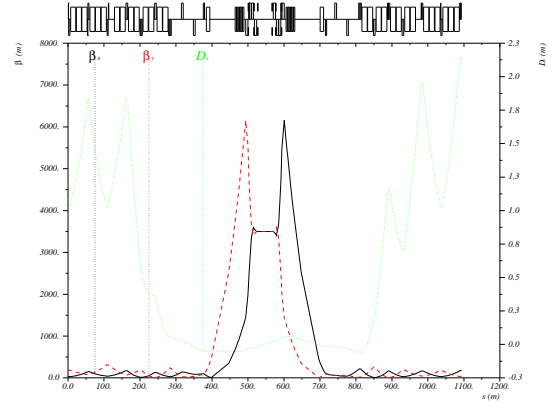} 
\end{turn}
\caption{ Very \hbt  optics with $\beta^{*}$=3500 m in Ring 1 
          around IP1, Version 6.0.}
\label{vhb-ir5v6}
 \end{center}
\end{figure}

A solution for measuring the total cross section in the vertical plane
and the Coulomb scattering in the horizontal plane with the Roman pots 
stations between  $Q6$ and $Q7$ ($RP3$ in the lower part of \ref{ir1}) 
could be found if $Q4$ is doubled in strength and  $Q8$ is exceeding 7.6\%.
Figure \ref{vhb-ir5v6} shows the solution for  $\beta^{*}$=3500 m in Ring 1
calculated with MAD9.
The most significant parameters for the total cross section and 
Coulomb experiments are summarized in tables \ref{ipd-ir5v6-3500} and 
\ref{ipd-ir5v6-3500-coul} respectively. From table 
\ref{ipd-ir5v6-3500-coul} we observed that $|x_{d}/\sigma_{x_d}|$ for nominal 
emittance is half of the required value to perform the measurement.
The problem could be solved by reducing the emittance by four, i.e the
emittance in the early days.

\begin{table}[htb]
\centering
\begin{tabular}{| l | c | c |} \hline
$\epsilon_{z}$     & \multicolumn{1}{c|}{$5.03\ \ 10^{-10}$} & m rad  \\ \hline
 
$\beta_{z}^{*}$    & \multicolumn{1}{c|}{3500.0} & m  \\  
$\alpha_{z}^{*}$   & \multicolumn{1}{c|}{0.0}    & \\  
$D_{x}^{*}$        & \multicolumn{1}{c|}{0.0}    &m  \\  
$D_{x}^{'*}$       & \multicolumn{1}{c|}{0.0}    & \\ 
$\sigma_{z}^{*}$   & \multicolumn{1}{c|}{1.33}   &mm  \\ 
$\sigma_{z}^{'*}$  & \multicolumn{1}{c|}{0.38}   & $\mu$rad  \\   \hline

\multicolumn{3}{|c|}{measurement vertical}  \\ \hline

\multicolumn{3}{|c|}{detector between $Q6-Q7$ }  \\ \hline
$\beta_{y_d}$         &  18.4  &m  \\  
$\Delta \mu_{y_d}$    &   0.750& 2$\pi$  \\ 
$M_{y,11_{d}}$        &   0.0  &   \\
$M_{y,12_{d}}$        &-253.4  &m  \\
$y_{d}$               &  -3.62 &mm  \\ 
$|y_{d}/\sigma_{y_d}|$&  37.7  &  \\
$|\theta_{y_{min}}|$  &   8.4  &$\mu$rad \\ 
$|t_{y_{min}}|$       &   0.003&GeV$^2$  \\ \hline

\end{tabular}
\caption{ Performance of a precise total cross section measurement 
   at the IP and at the detector place of Ring 1 
  for optics with $\beta^{*}$=3500, 
 Version 6.0 at 7 TeV for nominal emittance and the Roman Pots
 between $Q6$ and $Q7$.
  $ |\theta_{y_{min}}| = \sqrt{2} y_{d} / M_{y,12_{d}}$
  with  $y_{d}$ = 1.5 mm.}
\label{ipd-ir5v6-3500}
\end{table}

\begin{table}[htb]
\centering
\begin{tabular}{| l | c | c | c | } \hline

$\epsilon_{z}$     & \multicolumn{1}{c|}{$5.03\ \ 10^{-10}$} 
                   & \multicolumn{1}{c|}{$1.258\ \ 10^{-10}$} & m rad  \\ \hline
 
$\beta_{z}^{*}$    & \multicolumn{2}{c|}{3500.0} & m  \\  
$\alpha_{z}^{*}$   & \multicolumn{2}{c|}{0.0}    & \\  
$D_{x}^{*}$        & \multicolumn{2}{c|}{0.0}    &m  \\  
$D_{x}^{'*}$       & \multicolumn{2}{c|}{0.0}    & \\ \hline
$\sigma_{z}^{*}$   & \multicolumn{1}{c|}{1.33}   
                   & \multicolumn{1}{c|}{0.66}   &mm  \\ 
$\sigma_{z}^{'*}$  & \multicolumn{1}{c|}{0.38}   
                   & \multicolumn{1}{c|}{0.19}   & $\mu$rad  \\   \hline

\multicolumn{4}{|c|}{measurement horizontal}  \\  \hline
\multicolumn{4}{|c|}{detector between $Q6-Q7$} \\ \hline

$\beta_{x_d}$         &\multicolumn{2}{c|}{177.4}   &m  \\  
$\Delta \mu_{x_d}$    &\multicolumn{2}{c|}{0.250} & 2$\pi$  \\ 
$M_{x,11_{d}}$        &\multicolumn{2}{c|}{0.0}   &   \\
$M_{x,12_{d}}$        &\multicolumn{2}{c|}{787.9}   &m  \\ \hline
$x_{d}$               &  2.5 &  2.5   &mm  \\ 
$|x_{d}/\sigma_{x_d}|$&  8.4   & 16.8   &    \\
$|\theta_{x_{min}}|$  &   2.7   & 2.7   & $\mu$rad \\ 
$|t_{x_{min}}|$       &   0.0004& 0.0004   & GeV$^2$  \\ \hline

\end{tabular}
\caption{ Performance of a Coulomb measurement at the IP and at the 
  detector place of Ring 1 
  for optics with $\beta^{*}$=3500, 
 Version 6.0 at 7 TeV for different emittance values with the Roman Pots
 between $Q6$ and $Q7$.
  $ |\theta_{x_{min}}| = \sqrt{2} x_{d} / M_{x,12_{d}}$
  with  $x_{d}$ = 1.5 mm.}
\label{ipd-ir5v6-3500-coul}
\end{table}

\section{Detector acceptance}
\subsection{Total cross section}
\begin{figure}[htb]
  \begin{center}
    \leavevmode
     \includegraphics*[width=70.0mm]{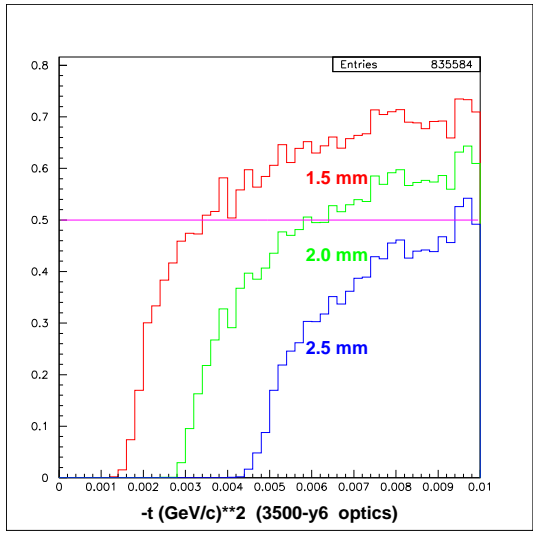} 
\caption{ Acceptance for total cross section.}
\label{dify6}
 \end{center}
\end{figure}

With the parameters of table \ref{ipd-ir5v6-3500}, 
we have  the results plotted in figure \ref{dify6}.
The three curves correspond to 15, 20 and 25 $\sigma_{y_d}$   
where $\sigma_{y_d} = 0.097$ mm is the vertical
beam size (rms) at the detector place, ($RP3$ in lower part of figure 
\ref{ir1}).
Adopting a conservative assumption (20  $\sigma_{y_d}$ = 2.0 mm) for
the approach distance, 
an efficiency better than $50 \%$ is reached at
$-t = 0.01$  GeV$^{2}$.

\subsection{Coulomb scattering}
\begin{figure}[htb]
  \begin{center}
    \leavevmode
     \includegraphics*[width=70.0mm]{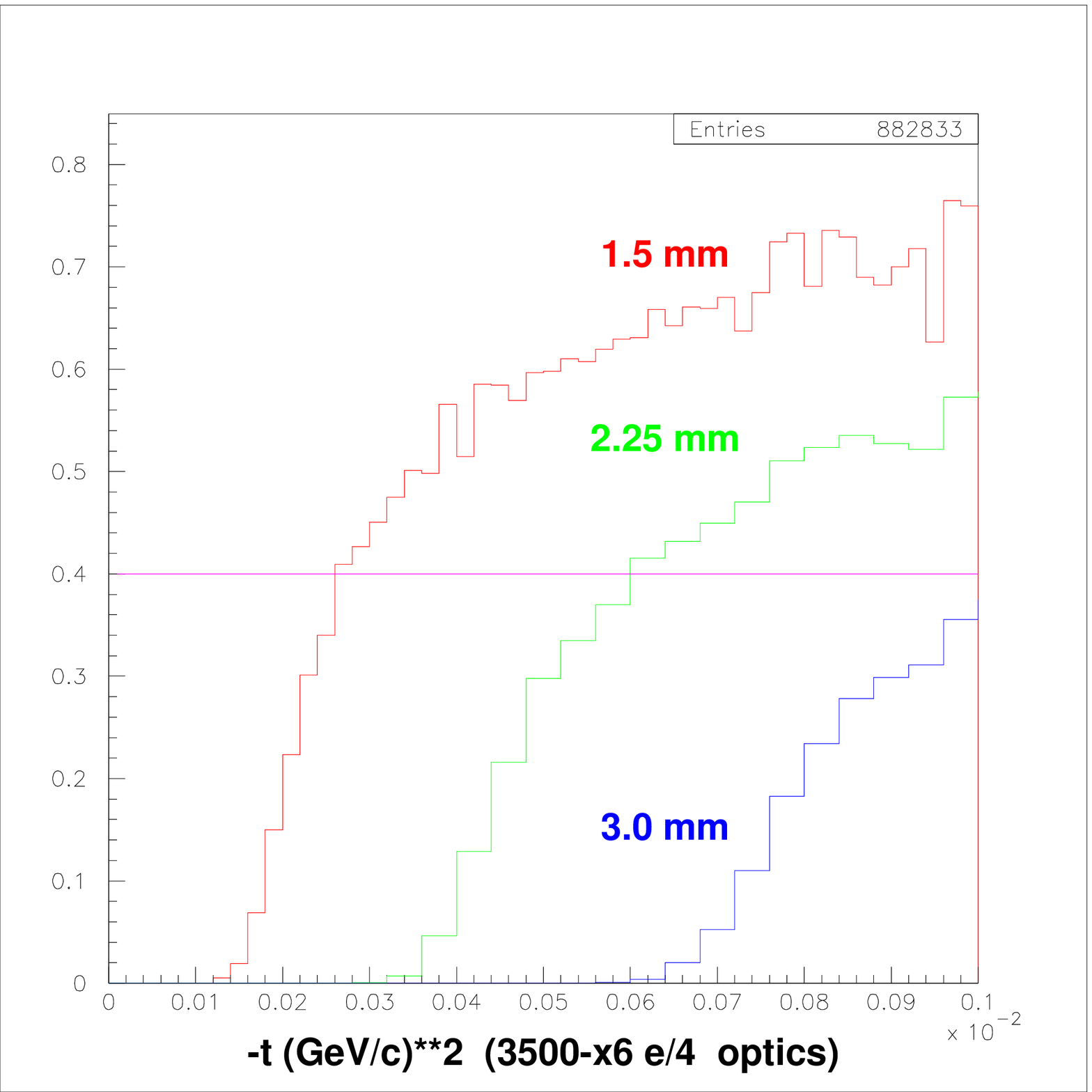} 
\caption{ Acceptance for Coulomb scattering.}
\label{difx6e4}
 \end{center}
\end{figure}

The crucial point for Coulomb scattering is to be able 
to reach  down to -t values up to  
$-t_{0}$ and beyond. Figure \ref{difx6e4} represents 
the  geometrical acceptance of a detector  2.0 mm x 2.5 mm
with the parameters of table \ref{ipd-ir5v6-3500-coul},
as a function of the minimal distance of approach  to the beam:  
10, 15 and 20 $\sigma_{x_d}$, where $\sigma_{x_d} = 0.149$ mm is the 
horizontal beam size (rms) at the detector place.

With a minimal approach distance of 2.2 mm (15  $\sigma_{x_d}$), 
an efficiency better than $40 \%$ is reached at
$-t = 6 \times 10^{-4}$  GeV$^{2}$.

\section{CONCLUSION}

A very \hbt optics ($\beta^{*}=3500m$), 
for a precise measurement of the total cross section and 
the Coulomb scattering at the LHC 
has been studied.  It requires some minor hardware modifications of the
present LHC set up.  With realistic assumptions as to the 
minimum beam  distance  approach, an acceptance 
good enough is obtained in both cases.

\end{document}